# TESTING AGILE REQUIREMENTS MODELS

**Jewgenij Botaschanjan, Markus Pister, Bernhard Rumpe**

**Software & Systems Engineering,
Technische Universität München,
Boltzmannstr. 3, 84758 Garching/Munich, Germany
http://www4.in.tum.de/**

**Abstract:**
This paper discusses a model-based approach to test software requirements in agile development processes. The use of models as central development artifact needs to be added to the portfolio of software engineering techniques, to further increase efficiency and flexibility of the development beginning already early in the requirements definition phase. Testing requirements is one of the most important techniques to give feedback and to increase the quality of the result. Therefore testing of artifacts should be introduced as early as possible, even in the requirements definition phase.

## 1 Introduction

One main purpose of requirements is to describe the functionality of software. Thus requirements often serve as a basis for contracts as well as for communication between customers and developers. However, they are usually captured in natural language accompanied by a few top-level informal drawings like use cases or activity diagrams that denote the structure of the functionalities in an abstract way. The disadvantage of natural language is that the developer has to cope with its ambiguity. The usage of precise or even formal descriptions for requirements helps to get along with this problem. However, preciseness of the language used does not mean that the description needs to be very detailed. Instead, given an appropriate language, one can specify very precise and very abstract. Further on precision also allows increasing the degree of tool support. Especially in innovative environments, where the requirements as well as the design and the implementation change rapidly, the probability for inconsistencies between formulated requirements and the implemented system is very high. Simulations of the behavior described in the specification and automatisms to synchronize the requirements with the design and the implementation increase as well quality of the result as efficiency of the development.

The approach taken here consists of a number of partially already well proven techniques, applied in different areas. The key idea is to combine the advantages of these techniques and apply them in at least one of their domains. The idea of upfront test design and refactoring comes from the Agile Methods Community, namely Extreme Programming, the use of modeling techniques from object-oriented software development methods (or from software engineering best practices in general), and the intensive use of code generators including behavior generators from automotive industry, where simulation and lately also production code generation from high level, state based modeling techniques is common already. The goal of this paper is to combine these concepts and transfer them to the already early phases of requirements modeling.

The usage of the Unified Modeling Language (UML, [1]) as a formal requirements description language is the main topic of Section 2. As primary technical element of model-based development, the form of model-based tests for the production code is discussed in Section 3. In Section 4 the validation of the requirements models is considered to ensure a high quality of the specification. As a remark for further work, the evolution (refactoring) of models is discussed in Section 5. Section 6 gives a conclusion.

## 2 Requirements Modeling with Executable UML

The UML undoubtedly has become the most popular modeling language for software intensive systems used today. Based on the ongoing standardization process its precision increases. Thus with some adaptations and interpretation guidelines for the language an unambiguous description can be created with relatively small effort. The UML consists of as many as nine kinds of diagrams usable for the description of the architecture and the design of the software. This variety of diagrams can be used for requirements models as well. Therewith the developers do not have to handle different and incompatible modeling languages within the same project.

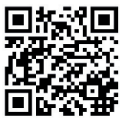

Though using the same language, there are some distinctions between design and requirements models:

1. Requirements models are usually less detailed and less precise than design models.
2. Requirements models describe properties of the system as a black box and do not describe the internal structure.
3. Requirements models often model the system and the environment (neighbor systems as well as user behavior), whereas design and implementation models concentrate on the system under development.

Regarding point 1, there is a general misconception about formality and precision of languages and statements: A requirement can and should be captured using a precise formal language. But even in a formal language it is possible to formulate abstract and thereby imprecise statements that only describe the details really needed and not anything more [9]. Although the UML has some deficits in allowing an abstract specificational modeling style, it is well suited to model abstract requirements.

Regarding point 2, it is clear today, that abstract behavioral specifications work for algorithms, such as sorting. In business information systems instead, we observe the data structure and accompanying functionality to be an integral part of the requirements model even so it serves also as part of the design. Thus most requirements engineering approaches distinguish between system requirements describing the black box view and constraints influencing the design, architecture and implementation [7]. One possibility to distinguish between data structures that are fixed by requirements / constraints and "auxiliary" data structures that only assist to describe required behavior and are therefore subject to change later, is to mark them explicitly, e.g. through stereotypes like «requirement» or «auxiliary data structure».

Requirements models are usually build from the user view of the system. Though having vague ideas of the design of the application the user often describes the system by formulating exemplary working steps that should be supported. By this, the software is described as a service to support working situations. These exemplary situations and the interaction with the software system can be described using sequence and object diagrams, but also using OCL and class diagrams for data structures and invariants.

The task of design models however is to define interfaces and precisely describe the behavior in a white box view of the system. The black-box description provided by requirements is refined into the "white-box" architecture and design of the system. This motivates the notion of refinement, defined as a mapping between an abstract system interface and a set of concrete design elements (e.g. interfaces or classes). In today's practice however, we have continuously evolving systems. Thus requirements, design and implementation evolve and the mapping between them has to be synchronized continuously.

This task needs tool and methodological support. Some UML-based tools today offer functionality to directly simulate models or generate at least parts of the test code for the software. As tool vendors work hard on continuous improvement of this feature, this means a sublanguage of UML will become a high-level programming language and modeling at this level becomes identical to programming. This raises a number of interesting questions, mainly dealing with the implications of using an executable UML in the development process. For example the degree of abstraction has an influence on the complexity of models. It is discussed, how fine grained an execu-

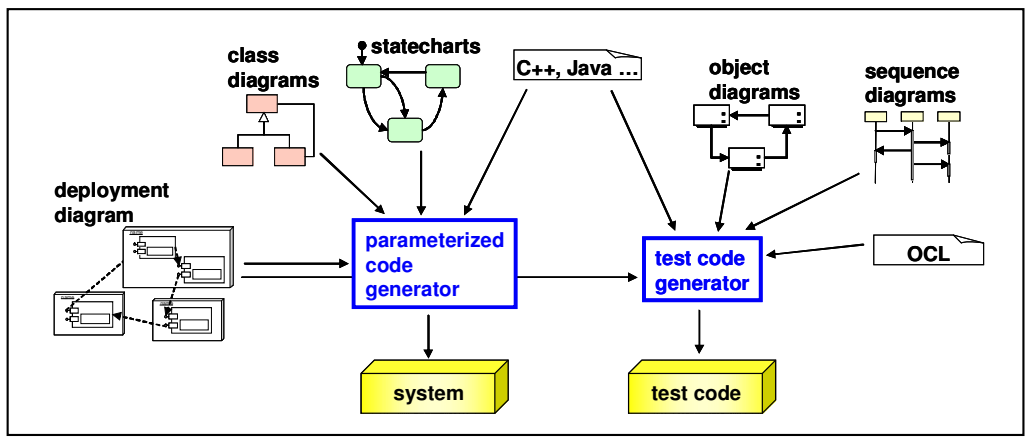

**Figure 1: Mapping of an UML –requirements model to code and tests.**

table model can be without loosing its clarity.

In [2,3] we have partly discussed issues like that and have demonstrated, how the UML in combination with Java may be used as a high-level programming language (see Fig. 1). Executable models are usually less abstract than design models, but they are more compact and abstract as the implementation. Therefore, having an executable modeling language for requirements definition is not a contradiction, but instead an important tool for analysis of requirements. Among others the UML can be used for modeling tests at various levels (class, integration, and system tests). Thus it can be used to describe tests for requirements at many abstraction levels.

## 3  Model-based Testing

The use of models for the definition of tests and production code can be manifold:

- Code or at least code frames stuffed with default behavior can be generated from a requirements model.
- Test cases can be derived from requirements models that are not used for constructive generation of production code. For example behavioral models, such as statecharts, can be used to derive test cases that cover states, transitions or even paths.
- Models can be used for an explicit description of a test case or a part thereof.

The first two uses are for example discussed in [11]. Since the nature of requirements models is mainly descriptive, this section concentrates on the test case generation from requirement models and the use of models to describe tests. There already exists a huge variety of testing strategies [10, 11]. A typical test, as shown in Fig. 2 consists of a description of the test data, the test driver and an oracle characterizing the desired test result. In object-oriented environments, the test data can be described by an object diagram (OD). It shows the necessary objects as well as concrete values for their attributes and the linking structure. The test driver can be modeled using a simple method call or, if more complex, a sequence diagram (SD). An SD has the considerable advantage that not only the triggering method calls can be described, but it is possible to model desired interactions and check object states during the test run.

For this purpose, the Object Constraint Language (OCL, [12]) is used to support the description of properties during and after the test run. It has proven efficient to model test oracles using a combination of object diagrams and OCL properties. An object diagram in this case serves as a fine grained property description and can therefore be rather incomplete, just focusing on the desired effects. The OCL constraints used can also be general invariants or specific property descriptions. The advantage of using OCL for the property description is the possibility to formulate general predicates for values of attributes. Thus, the same result description or at least parts of it can be reused for several test inputs.

As already mentioned, being able to use the same, coherent language to model the production system and the tests gives us a good integration between both tasks. It allows the developer to immediately define tests for the constructive model developed. It is imaginable that in a kind of "test-first modeling approach" (see [4, 5]) the test data in form of possible object structures are developed before the current implementation. This test-first approach perfectly fits with the modeling of requirements. The models mostly describe the behavior of the system at the interfaces in a form that these requirements can be used as test drivers as well. For the creation of tests the developers only have to create representative test inputs. The following section explores the issue of describing behavior in greater detail.

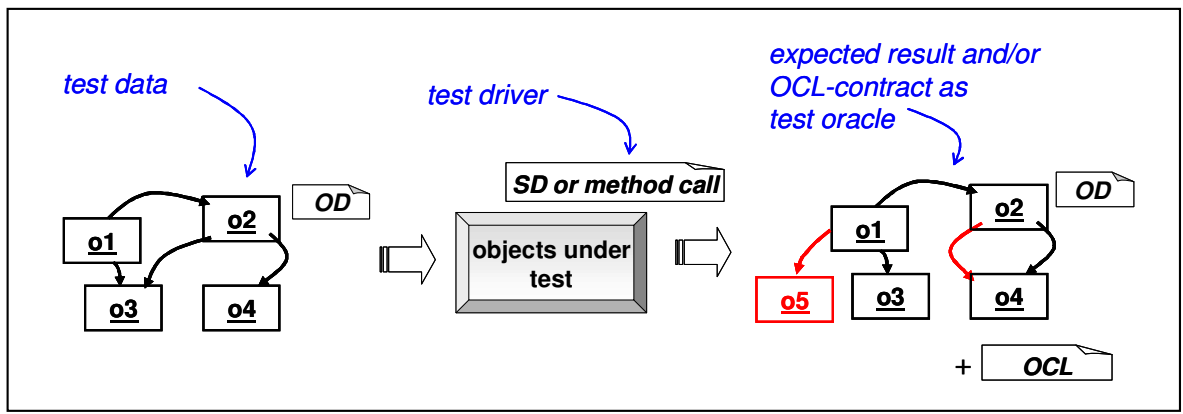

**Figure 2: Structure of a Test modeled with Object Diagrams (OD) and Sequence Diagram**s

## 4 Testing Requirements

As explained in Section 2, at least functional requirements can be defined using the UML. Based on an "architecture" model in form of one or more class diagrams, requirement definitions may also have constraints formulated in OCL, state models of individual elements in the system denoted as statecharts and exemplary descriptions in form of sequence and object diagrams. Figure 3 shows such a sequence diagram that is typically used for a test case validation. The initial part (marked with the steroptype «trigger») acts as driver for the test, the other interactions of the sequence diagram are observations to be made during a successful test run. This sequence diagram was derived almost directly from a requirements definition and only little extra effort was necessary to transform it into a test (namely adding the stereotype). The underlying data structure was reused from exemplaric requirements models, namely an object diagram. Furthermore the OCL property description at the end of the sequence diagram is perfectly suited for a post-condition check.

Additionally to the validation of the implementation to ensure the conformity with the modeled requirements, the quality of the requirement definitions themselves has to be validated. It is of strong interest to gain feedback on the formulated requirements as early as possible to prevent long lasting and therefore expensive errors. Besides through reviews, the primary technique for that purpose is the animation of the requirements models and furthermore the run of automated tests against the animated models.

As explained, sequence and object diagrams are perfectly suited to define tests, assumed they are detailed enough. This imposes some additional work on the requirements model, but results in a highly valuable early feedback already in the requirements definition phase. The major problem with this approach is usually the missing complete behavioral description for elements that participate in a test. For example it may be that the behavior of a component is only defined in terms of a finite number of sequence diagrams, but no implementation or statechart is given. In this case, the implementation has to be simulated according to the given information. We can for example use the approach from Ingolf Krüger [13] or the Play-In/Play-Out Approach [14] to construct the overall behavior from the given sequence diagrams.

This technique is for example useful for components of the environment or neighbor systems that actively participate in tests, but will not become a part of the implementation. It allows to animate the environment for testing purposes, but not to test the environment itself, because if tests and production code are generated from the same models, both tests and code are consistent with each other and therefore errors in the models cannot be detected.

The idea of automatically creating an executable simulation from the requirements model is an extension of the concept of rapid prototyping. The automatic generation delivers immediately a prototype that demonstrates the behavior of the system. By receiving this fast feedback, the developers and customers can early get an imagination of the running system and adapt the specification before having designed the implementation. Additionally to checking the consistency, this technique can especially be used for the validation of the completeness of a specification by presenting the customer the whole defined functionality. This decreases the effort to be spent for changes and therefore leads to a higher quality of the software and to reduced development costs.

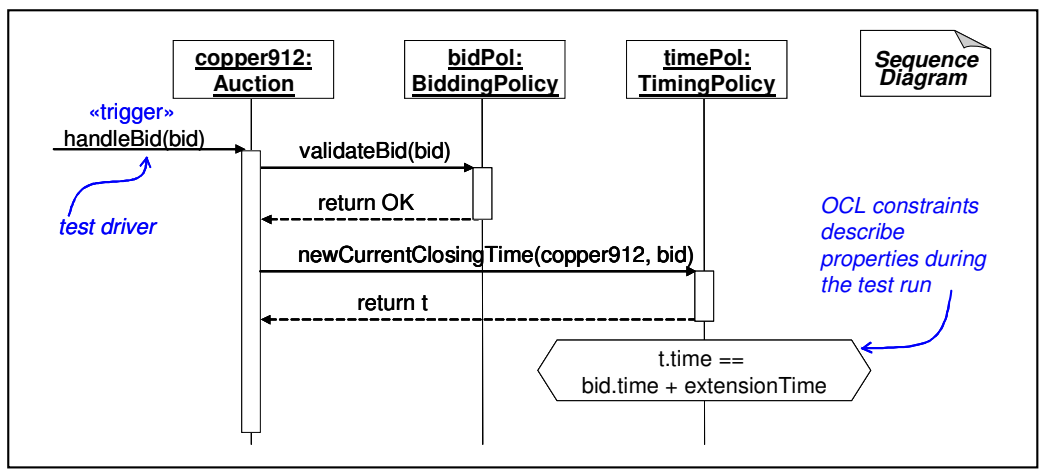

**Figure 3: Test Case Description with Sequence Diagrams**

However, the generation of automated tests goes considerably beyond the mere prototyping approach. First, the automated tests can be rerun and used in regression testing not only during the requirements modeling, but also during design and implementation. Second, automated tests can be run by everyone in the project, not only by the experts who know the requirements. Third, experiments with the testing approach have shown that in the long run automated tests pay off, even though it is initially more time consuming to develop tests. Using requirements models for that purpose, however, also increases efficiency in test development.

Of course, not all inconsistencies can be detected within the requirements model and modifying or extending the models over time is necessary as well in requirement as in design models. Therefore the next section outlines the evolution of models.

One might argue that requirements elicitation is nowadays usually a step where very informal techniques are used. Mostly, requirements elicitation results in a specification are written in natural language. Only after its approval the requirements are mapped into high level analysis models. However, we would like to stress again that we strongly favor the approach taken here, where the understood requirements are translated into a machine understandable, ideally also executable form. This greatly eases the feedback of the elicitation process with the user, allows automated consistency checks and test, as well as analysis for completeness of requirements. This, however, is not applicable in every project, as there may be obstacles like user demands, legal requirements, or very large groups of project members. But we are sure that an increasingly higher portion of projects will be able to use an approach that includes the concepts described here.

## 5  Model Evolution using Automated Tests

Neither requirements nor design models are initially correct. It is expected that the development and maintenance process is capable of being flexible enough to dynamically react on changing requirements. In particular, enhanced business logic or additional functionality should be added rapidly to existing systems, without necessarily undergo a major re-development or re-engineering phase. The extension of functionality of the software often requires a transformation of design as a preliminary step. The refactoring techniques for Java [6] have shown that a comprehensible set of small and systematically applicable transformation rules seems optimal. Transformations, however, cannot only be applied to code, but to any kind of model. Figure 4 shows a transformation on a class diagram. It essentially moves up two elements along a hierarchy. This may e.g. happen when it becomes apparent in the requirements elicitation process that a data structure element is common to several subclasses or that an operation behaves uniformly on several branches of the inheritance tree. Even if this transformation looks rather easy, it might trigger a number of additional refactorings to ensure that the specifications resp. implementations for the moved method are consistent. A number of possible applications for refactoring are discussed in [8].

Many of the necessary transformation steps are rather simple and easy to apply. However, some transformations are deeper changes on the system and their application involves context conditions. Fortunately most of them can be structured in a number of individual steps. This means that the power of these simple and manageable transformation steps comes from the possibility to combine them and evolve complex designs in a systematic and traceable way.

Following the definition of refactoring [6], transformational steps for structure enhancement do not affect "externally visible behavior". By "externally visible behavior" Fowler in [6] basically refers to behavioral changes visible to the user. However the change of requirements mirrors explicitly in a change of the external behavior. The difference between changes in design and requirements can be precisely understood by introducing an abstract "system border". This border serves as interface to the user, but may also act as interface to other systems. It is therefore necessary to explicitly describe, which kind of behavior is regarded as externally visible. For this purpose tests are the appropriate technique to describe behavior, because tests are already available through the development process and tests are automated which allows us to check the effect of a transformation through inexpensive, automated regression testing. A test case created from a requirements model thus acts as an "observer" of the behavior of a system under a certain condition A test can thus be understood as "implicitly defining an observed system border", like shown in Figure 5.

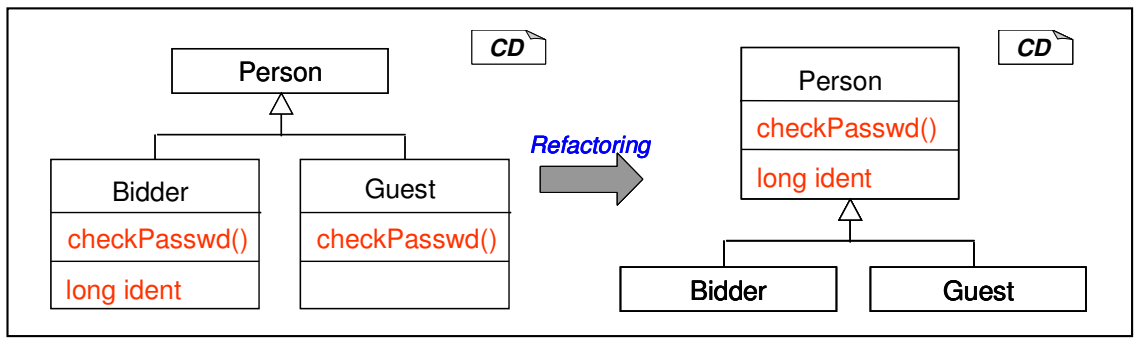

**Figure 4: Example for a Model Transformation**

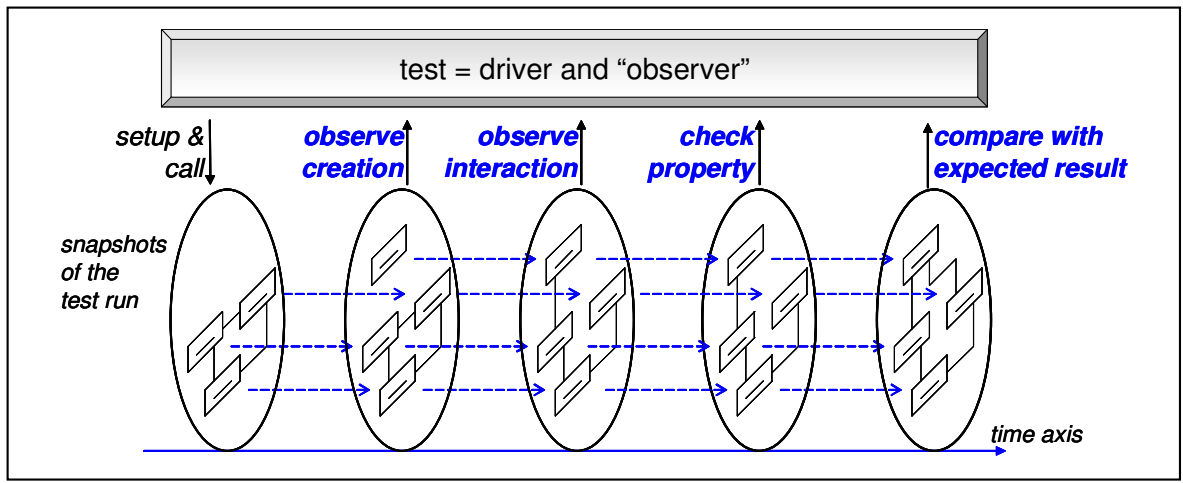

**Figure 5: Tests implicitely defining the System Border**

Depending on the level and origin of tests, they may have a very different observational interface. To be able to manage tests in case of evolution of requirements or implementation, it is therefore inevitable to test against clear interfaces (namely "published", stable ones). A number of test patterns like defined in [2] helps to define tests in a rather clear way. However, it surely needs also a better understanding of how to cut tests from requirements, how deep these tests should go and how stable requirement based tests are in the evolutionary development process. For that purpose we need improved test generation tools and in particular more empirical assessments in real projects.

## 6  Conclusions

The discussion in this paper can be summarized as a pragmatic approach to introducing feedback in the requirements definition through model-based testing. It suggests using models as primary artifact for requirements and design documentation, code generation and test case development. The validation of the requirements by executable UML models allows an early detection of errors in the specification. The test case generation from requirements models also allows checking the correctness of the implementation and thus enables to validate the correctness of transformations of the code. Finally the management of evolution of the requirements, the design and the implementation is systematically supported by a set of predefined transformations on the models.

However, the methodology sketched here still is a proposal. Some major efforts still need to be done and are currently under development. On the one hand, even a number of works on various model transformations do exist. But, they are not very well put in context and not very well integrated with the UML in its current version.

On the other hand, model based evolution will become successful only if well assisted by tools. This includes parameterized code generators for the system as well as for executable test drivers, analysis tools and comfortable help for systematic transformations on models.

**Reference**

1. OMG. Unified Modeling Language Specification. V1.5. 2002.
2. Rumpe, B. Agiles Modellieren mit der UML. Habilitation Thesis. Technische Universität München, Institut für Informatik, 2003.
3. Rumpe, B. Executable Modeling with UML. A Vision or a Nightmare? In: Issues & Trends of Information Technology Management in Contemporary Associations, Seattle. Idea Group Publishing, Hershey, London, pp. 697-701. 2002.
4. Link J., Fröhlich P. Unit Tests mit Java. Der Test-First-Ansatz. dpunkt.verlag, 2002.

5. Beck K. Aim, Fire (Column on the Test-First Approach). IEEE Software, 2001.
6. Fowler M. Refactoring. Addison-Wesley. 1999.
7. Wiegers Karl E., Software Requirements. Microsoft Press, 2003
8. Philipps J., Rumpe B.. Refactoring of Programs and Specifications. In: Practical foundations of business and system specifications. H.Kilov and K.Baclawski (Eds.), 281-297, Kluwer Academic Publishers, 2003.
9. Kiczales G., Lamping J., Mendhekar A., Maeda C., Lopez C., Loingtier J.-M., Irwin J. Aspect-Oriented Programming. In ECOOP'97 - Object Oriented Programming, 11th European Conference, Jyväskylä, Finnland, LNCS 1241. Springer Verlag, 1997.
10. Binder R. Testing Object-Oriented Systems. Models, Patterns, and Tools. Addison-Wesley, 1999.
11. Briand L. and Labiche Y. A UML-based Approach to System Testing. In M. Gogolla and C. Kobryn (eds): «UML» - The Unified Modeling Language, 4th Intl. Conference, pages 194-208, LNCS 2185. Springer, 2001.
12. Warmer J., Kleppe A. The Object Constraint Language. Addison-Wesley. 1998.
13. Krüger, I. Distributed System Design with Message Sequence Charts, Ph.D. Thesis, Technische Universität München, 2000.
14. Harel, D. and Rami, M.: Specifying and executing behavioral requirements: the play-in/play-out approach. In: Journal on Software and Systems Modeling, SOSYM. Vol:2(2), pg. 82 – 107, Springer-Verlag Heidelberg, 2003.